\documentstyle[12pt,epsfig]{article}

\begin{document}

\title{ Bursts and Shocks in a Continuum Shell Model}
\author{ K. H. Andersen, T. Bohr, M. H. Jensen and P. Olesen \\
The Niels Bohr Institute\\
Blegdamsvej 17\\
DK-2100 Copenhagen {\O}\\
DENMARK}

\maketitle

\begin{abstract}

We study a ``burst" event, i. e. the evolution of an initial condition 
having support only in a finite interval of $k$-space, in the 
continuum shell model due to Parisi 
\cite{Parisi}. We show that the continuum equation without forcing 
or dissipation can be explicitly written in characteristic form and that 
the right and left moving parts can be solved exactly. When this is 
supplemented by the appropriate shock condition it is possible to find 
the asymptotic form of the burst.
\vspace*{3cm}
\begin{center}
Contribution to the proceedings of the Conference: \\
{\it Disorder and Chaos},\\
in honour of {\bf Giovanni Paladin},\\ September 22-24, 1997 in Rome.
\end{center}

\end{abstract} 
\clearpage
%\newpage

\section{Introduction}

In the study of fully developed turbulence the so-called shell models 
have become an important tool. In these models the essential 
ingredients are the conservation laws and the assumption of locality 
on a discrete set of exponentially growing momentum shells. The 
models are constructed such that Kolmogorov's 1941 scaling law is 
built in as a fixed point in the absence of forcing and viscosity.

It has recently been found (see e.g. \cite{book} and references 
therein) that these models actually capture important features of the 
strongly intermittent behaviour seen in experiments, which gives rise 
to corrections to the simple Kolmogorov dimensional scaling. In 
order to understand how such corrections come about, it is necessary 
to study the strong ``burst" events which destroy the homogeneity of, 
say, the dissipation field and presumably lead to the appearance of 
multiscaling.

In 1990 Parisi \cite{Parisi} suggested a continuum 
approximation of a shell model and showed that solitary wave 
excitations would exist on top of the Kolmogorov spectrum
(a short presentation of his approach can be 
found in \cite{book}). It has, 
however, recently been stressed \cite{Fridolin} that the shell models 
have another (trivial) fixed point, where the fluid is quiescent, and 
that this state seems to be approached before a large burst. In the 
present paper we shall thus study how a single pulse (a disturbance 
initially localized in wave number space) will propagate into a 
motionless fluid. The details of how such a burst propagates are quite 
interesting and unexpected and although it remains to be seen, 
whether the shocks which are generated represent the bursts in real 
turbulent fluids, we believe that they are worth discussing since the 
methods that we use might be generalized to more realistic models. 
We feel sure that Giovanni Paladin would have enjoyed our story -- 
indeed he was the one who introduced us to the shell models and later 
contributed and inspired so much of the work in this field.

\section{The GOY shell model and its continuum limit}

%\subsection{Shell models}
The basic idea behind the shell model is that the Navier-Stokes 
equation,
when considered in Fourier space, receives contributions to the time 
derivative
of the velocity from the velocities in a triangle of $k$-vectors. This
condition is mimicked by taking 
$k$-vectors as the one-dimensional discrete set (``shells") 
$k_n = k_0 r^n$, and introducing only a single complex field $u_n$ 
for each $k$ with
interactions only between nearest and next nearest neighbours. The 
field $u_n$ represents the characteristic velocity differences across 
the $n$th shell, and, in the limit of vanishing viscosity and forcing, 
energy conservation is imposed in the form
\begin{equation}
\sum |u_n|^2 = {\rm const}
\end{equation} 

The most studied model is the so-called GOY model \cite{book} (after 
Gledzer, Okhitani and Yamada)  
\begin{equation}
({\frac{du_n}{dt}} + \nu k_n^2 u_n)^* = - i k_n ( u_{n+1} 
u_{n+2}
-{\frac{\delta}{r}}u_{n-1} u_{n+1}-{\frac{1-\delta}{r^2}}u_{n-1} 
u_{n-2}) + F_n
\end{equation}
where $\nu$ is the viscosity and $F_n$ is the forcing. $\delta$ is a
free parameter, but becomes fixed at $\delta=1/2$ when the conservation
of helicity is also satisfied.

\subsection{The Parisi continuum limit}

It is an interesting question how to define the shell model in the limit
where the $k$-variable becomes continuous. A particular way of 
reaching this 
limit is to take $r\rightarrow 1$: By writing the distance between the 
shells 
as $r=1+\epsilon$ with $\epsilon\ll 1$, we have $k_n\approx \exp 
(n\epsilon)$,
so with $n\sim$ const.$/\epsilon$ a continuous range of values is 
obtained 
for the variable $k$. 

To proceed we use a Taylor expansion of the type
\begin{eqnarray}
u_{n+1}(t)&=&u(k_{n+1},t)=u(\ln k_{n+1},t)\approx u(n 
\ln(1+\epsilon),t)+
\ln (1+\epsilon)\frac{\partial u_n}{\partial \ln k}\nonumber \\
&\approx& u(k,t)+\epsilon k\frac{\partial u(k,t)}{\partial 
k}\nonumber \\
&=&u(k,t)+\epsilon k\frac{\partial u(k,t)}{\partial k},
~{\rm with}~u(k,t)\equiv u(k_n,t)=u_n(t),
\end{eqnarray}
and similarly for $u_{n+2}, u_{n-1},$ and $u_{n-2}$.
To first order one obtains \cite{Parisi}:
\begin{equation}
 u_t^*+\nu k^2 u^*=
  -ik(2-\delta)\left(u^2+3ku u_k \right),
\label{firstorder}
\end{equation}
where $u_k \equiv \partial u /\partial k$ etc.
The higher order corrections to (\ref{firstorder}) and the 
convergence to the discrete model will be examined in 
\cite{longarticle}.
By rescaling time with $2-\delta$ we get the Parisi equation, which
in its inviscid, unforced form is:
\begin{equation}
  u_t^* + 3 i k^2 u u_k = - i k u^2
  \label{Parisi}
\end{equation}
In terms of the real and imaginary part ($u = a + i b$) it can be 
written as:
\begin{equation}
\label{a-Parisi}
a_t -3k^2 a b_k - 3k^2 a b_k = 2 k a b
\end{equation}
and
\begin{equation}
\label{b-Parisi}
b_t - 3k^2 a a_k + 3k^2 b b_k = k (a^2 - b^2).
\end{equation}
For the case of $a=0$,  a number of exact solutions to (\ref{b-Parisi}) have 
been found in \cite{olesen}.
\section{Solution of the inviscid Parisi equation}
We shall now study the inviscid, unforced Parisi equation 
(\ref{Parisi}) in the special case, where the initial condition is a 
single ``burst", i.e. that the field $u$ is only nonzero in some finite 
interval in $k$-space. 
\subsection{Direct simulation}
The Parisi-equation (\ref{Parisi}) or (\ref{a-Parisi})-(\ref{b-Parisi}) 
is a hyperbolic equation \cite{Whitham}, and hyperbolic equations are 
notoriously hard to solve
numerically due to the appearance of shocks in
the solution. These can be dealt with in a crude way by using upwind
differencing of the flux terms \cite{leveque}, but to do this, one need 
to figure out in which direction a shock is moving. For the real part 
(\ref{a-Parisi}) there is a minus sign on the flux term, signifying that 
a shock moves to the left (provided
both real and imaginary parts are positive). The imaginary part 
(\ref{b-Parisi}) is not so simple, since the flux term is composed of 
both a right and a left
moving part. Never the less upwind differencing works there too, by 
assuming that the dominant direction is to the right.

\begin{figure}[htbp]
  \epsfig{file=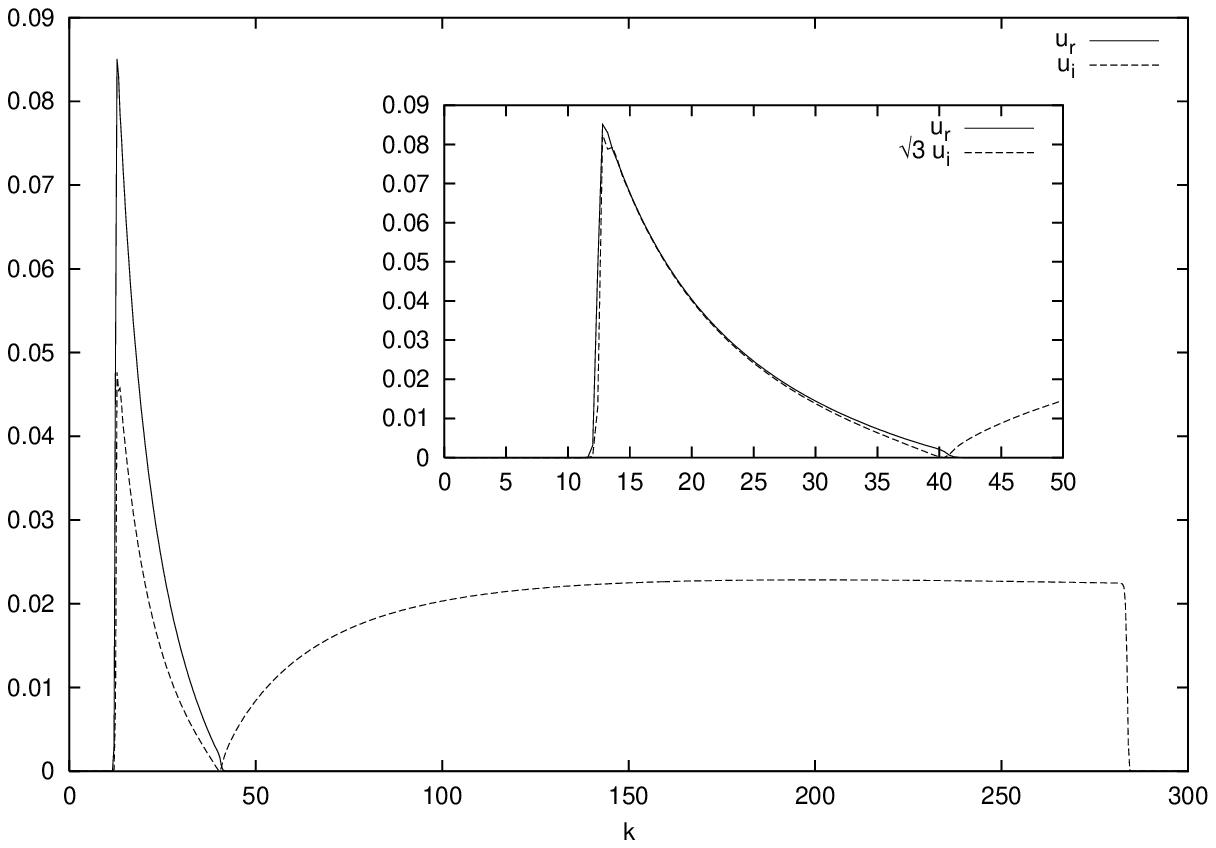, width=0.75\linewidth}
  \epsfig{file=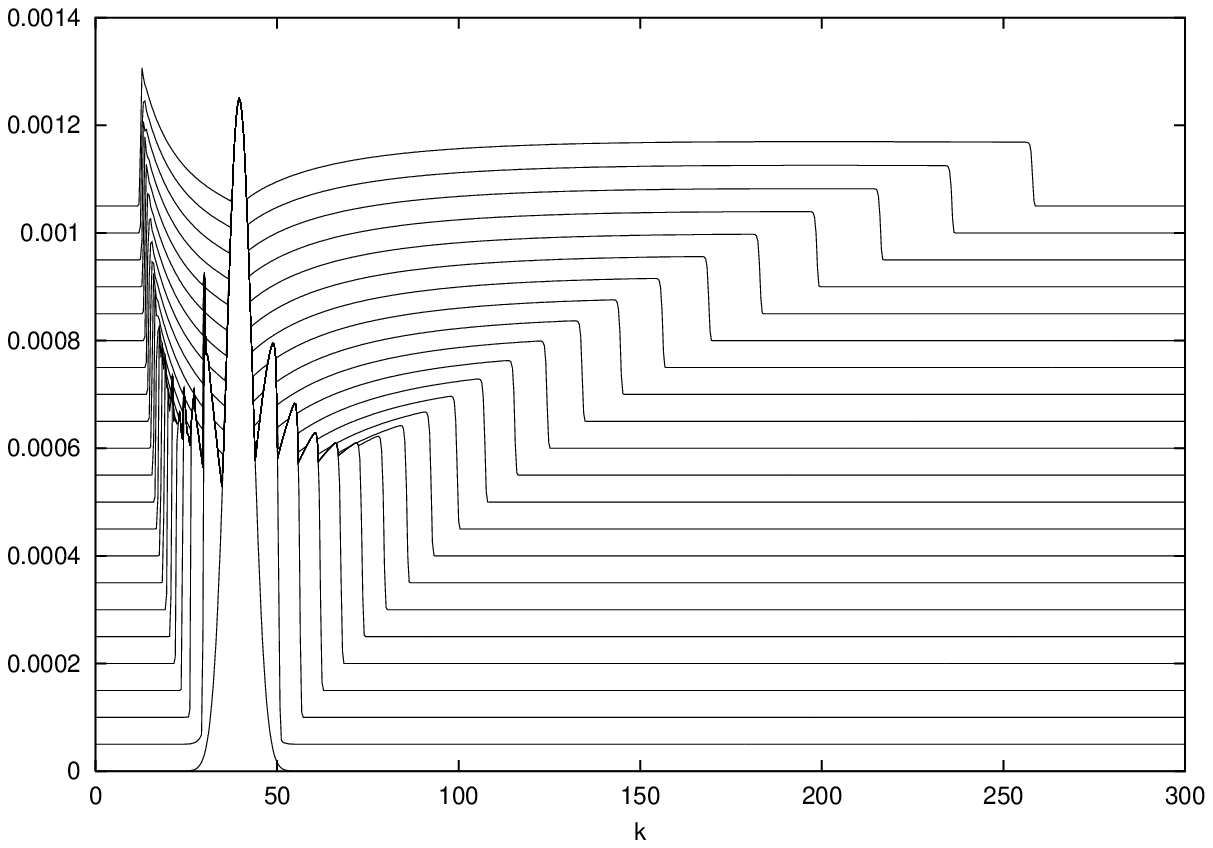, width=0.75\linewidth}
  \caption{A numerical solution of the Parisi equations, without
    forcing and viscosity. The upper figure shows the solution 
at $t = 10^{-3}$ after
    having been started with a gaussian peak at
    $k=40$. The inset on the uppermost figure shows the part moving 
to
    the left, and illustrate that the relation $a=\protect\sqrt{3} b$
    is fulfilled to a high 
    accuracy. In the lower figure only the imaginary part is shown in
    a space-time plot. Here the splitting of the initial pulse into a
    left- and right-moving part is clearly visible. The initial phase is $\pi/4$.}
  \label{fig:num1}
\end{figure}

An example of a numerical solution with semi-implicit update in time, 
is seen in figure \ref{fig:num1}. The initial condition is the zero
state, with a gaussian peak $a(k, t=0) = b(k,t=0) =
C \exp(-(k-k_0)^2)$,
with $k_0 = 40$ and $C = 0.25$. In order to neglect viscosity, the 
inertial subrange must be to the right of the initial perturbation.

A short while after the start, shocks are formed in
both directions. As was seen from the equations, the real part is
only able
to move to the left, while the imaginary part moves both ways. 
Thus only the imaginary part is important in the inertial
 subrange.
Setting $a = 0$ in the original equations, 
we get just a single real equation for the inertial range:
\begin{equation}
b_t + 3 k^2 b b_k= - k b^2 
\label{equ:rpulse}
\end{equation}
For the left-moving part of the pulse, 
the real and the imaginary parts are seen to become proportional. 
Inserting $b = Ca$ into the Parisi equation gives us two possibilities: 
$a=0$
(which we already treated) and $a = \pm \sqrt{3} b$. For  $a =
\sqrt{3} b$ the
equation for the imaginary part becomes:
\begin{equation}
b_t - 6 k^2 b b_k = 2 k b^2.
  \label{equ:lpulse}
\end{equation}
The explanation for the splitting of the pulse in this way will be given 
in the next section by writing (\ref{Parisi}) in characteristic form.

Another problem occurring when solving hyperbolic differential 
equations is to
find the correct position and velocity of the shocks. Strictly speaking, 
the
equations are not well-defined after the occurrence of a shock, because 
the solutions become
multivalued. To solve past the time when the shocks first appear one
has two choices: 1) to derive a shock condition by imposing 
conserved quantities (note that energy conservation alone
is not enough, since two conditions are needed) or 2) include a higher order
terms to smooth out the shock. When solving differential equations 
with finite
differences, there is always a small amount of ``numerical
diffusion'', corresponding to a $\partial^2 u/\partial k^2$ term
in the equations. This is not necessarily the correct higher order 
term,
and therefore the velocity of the shocks as seen in figure
\ref{fig:num1} is not necessarily correct. 
We shall show later how to 
obtain the correct shock conditions by imposing energy conservation, 
and thereby the correct asymptotics for the solution.

\subsection{Characteristic form for the Parisi equation}
Surprisingly, the ``source" term (i.e. the right hand side) of 
equation (\ref{Parisi}) can be removed by the simple substitution $u 
= k^{-1/3} v$ and $k = (2x)^{-3/2}$. This turns (\ref{Parisi}) into 
a complex equation reminding strongly of the Burgers equation
\begin{equation}
v_t^* -  i  v v_x = 0.
\label{v-Parisi}
\end{equation}
In terms of real and imaginary parts of $v$ (again denoted by $a$ 
and $b$) we get
\begin{equation}
\label{aa-Parisi}
a_t + b a_x + a b_x =0
\end{equation}
and
\begin{equation}
\label{bb-Parisi}
b_t + a a_x  -b b_x =0
\end{equation}
The calculations turns out to become simpler if (\ref{v-Parisi}) is
written in terms of modulus and phase, i.e.
\begin{equation}
\label{polar}
v = r e^{i \phi}
\end{equation}
whereby we find the equivalent equations
\begin{equation}
\label{r-Parisi}
r_t + r r_x\sin\theta  + {\frac{1}{3}}r^2 \theta_x\cos\theta =0
\end{equation}
and
\begin{equation}
\label{theta-Parisi}
\theta_t + 3 r_x  \cos\theta - r \theta_x \sin\theta =0
\end{equation}
where $\theta = 3 \phi$. This
can be written as a matrix equation
\begin{equation}
\label{hyp}
 {\bf q}_t + {\bf A} {\bf q}_x =0
\end{equation}
where ${\bf q}= (r,\theta)$ and
\begin{equation}
{\bf A} = \left( \begin{array}{cl}
r \sin\theta & {\frac{1}{3}}r^2 \cos\theta \\
3 \cos\theta &  - r \sin\theta
\end{array} \right)
\end{equation}
The characteristics are the eigenvalues of ${\bf A}$, which are
$\lambda = \pm r$ and the Riemann invariants can be found through 
the
left eigenvectors of ${\bf A}$ (or the right eigenvectors of the
transpose of ${\bf A}$ \cite{Chorin}). These eigenvectors are of the 
form
${\bf e}=(\alpha,\beta)$, and satisfy
\begin{equation}
\label{eigenv}
\beta = r \left ({\frac{\pm 1 - \sin \theta}{3 \cos \theta}}\right) \alpha
\end{equation}
The Riemann invariants satisfy
\begin{equation}
\label{r-Riemann}
{\frac{\partial J_{\pm}}{\partial r}} = \alpha
\end{equation}
and 
\begin{equation}
\label{theta-Riemann}
{\frac{\partial J_{\pm}}{\partial \theta}} = \beta
\end{equation}
and the Parisi equation (\ref{v-Parisi}) can be reformulated as the 
two equations
\begin{equation}
J_{\pm} = {\rm const} \,\,\,\, {\rm on \, the \, curve} 
\,\,\,\, {\frac{dx}{dt}}=\pm r(x,t)
\end{equation}
Only the ratio $\beta / \alpha$ is uniquely defined. One solution for 
(\ref{r-Riemann})-(\ref{theta-Riemann}) is
obtained by taking $\alpha = 1/r$ and thus
\begin{equation}
\alpha = {\frac{\partial J_{\pm}}{\partial r}} = 
{\frac{1}{r}}
\end{equation}
and
\begin{equation}
\beta = {\frac{\partial J_{\pm}}{\partial \theta}} =
{\frac{\pm 1 - \sin \theta}{3 \cos \theta}}
\end{equation}
which gives
\begin{equation}
J_{\pm} = \log(r (1\pm \sin \theta)^{1/3})
\end{equation}
Now, the Riemann invariants are only defined by their invariance 
along the characteristics, and thus we can equally well take them to 
be defined without the
logarithm, i.e.
\begin{equation}
\label{J}
J_{\pm} = r (1\pm \sin \theta)^{1/3}
\end{equation}
which of course is constant on the same locus. 

To complete the solution we can express $r$ and $\theta$ through 
$J_{\pm}$. 
We find
\begin{equation}
\label{r(J)}
r= \left({\frac{J_{+}^3 + J_{-}^3}{2}}\right)^{1/3}
\end{equation}
and
\begin{equation}
\label{theta(J)}
\sin\theta = {\frac{J_{+}^3 - J_{-}^3}{J_{+}^3 + J_{-}^3}}
\end{equation}

The formulation in terms of characteristics can now be used to 
understand the
splitting of the pulse found in section 3.1. The important point is that, 
since $r$ is nonnegative, one family of characteristics ($J_+$) can 
never move to the left
while the other one ($J_-$) can never move to the right. If the initial
condition has support in a limited region of $x$, say $[x_-,x_+]$ the 
same is
true of $J_\pm$. They both vanish (in the initial condition) outside of 
this
interval. For times $t>0$ we compute the field values by finding 
where, on the
$x$-axis (i.e. $t=0$) the characteristics going through the point 
$(x,t)$
emanate. Now, if $x>x_+$, the $J_-$-characteristic going through 
this point must
emanate from some $x_0 > x_+$ and thus $J_-(x,t)= J_-(x_0,0) 
=0$. This
means that either $r=0$ (which makes the entire field vanish) or 
$\sin \theta  =
1$ which means that $\theta$ has the constant value $\pi/2$ or $\phi = 
\pi/6$.
This corresponds exactly to the result of last section: that $a = 
\sqrt{3} b$
for the left moving pulse (which is moving right in the $x$-variable).

A completely similar argument, valid for $x< x_-$, shows that in this 
case $\sin
\theta =-1$ or $\theta=3 \pi / 2$ and $\phi = \pi/2$, which implies 
that $a=0$,
again in agreement with the result of last section for the right moving 
pulse. This shows that the regions outside $[x_-,x_+]$ are so-called simple wave
regions \cite{Whitham}.

When we said that $\theta = \pi/2 \Rightarrow \phi = \pi/6$ it is not
entirely correct. The value $\phi = \pi/6$ is only one possibility, the two 
others being $\phi = 5 \pi/6$ and $\phi = 3 \pi/2$. Likewise, $\theta = 3\pi/2$
can mean $\phi =  \pi/2$, $\phi = 7 \pi/6$ or $\phi = 11 \pi/6$. If
the initial phase is in the interval between two of these six values of $\phi$
corresponding, consecutively, to $\sin \theta$ having the value $+1$ and
$-1$, the values of $\phi$ selected will be precisely those two. The values
$\phi =\pi/6$ and $\phi = \pi/2$ given above correspond to an initial phase 
around $\pi/4$.
\subsection{Asymptotic analysis of the burst}
The splitting of the pulse into a right-moving part with $a=0$ and a 
left-moving part with $a=\sqrt{3} b$ makes it possible to give a 
complete solution for a single burst event.
We shall divide the analysis into the two simplified cases; forward
and backward moving parts. In each regime the field is described by a
single scalar equation. It is therefore possible to find the correct
shock condition using a single conservation law: energy conservation.
We have chosen the case
where the initial phase is around $\pi/4$, but the other cases can be
handled similarly.
\subsubsection{Forward propagation}
When the real part $a=0$ the equation for the imaginary part $b$ 
(which
we shall call $u$ in this section) is
\begin{equation}
\label{i-Parisi}
u_t + 3 k^2 u u_k  = -k u^2
\end{equation}
where we, for the sake of explicitness, use the original form instead of 
the ``Burgers form" (\ref{v-Parisi}).
Again, this is a hyperbolic equation, which can be 
solved by the method of characteristics. 
The characteristic equations are 
\begin{equation}
\label{char1}
k'(t) = 3 k^2 u
\end{equation}
\begin{equation}
\label{char2}
u'(t) = -k u^2
\end{equation}
and they can be solved e.g. if initial conditions are specified in the 
form 
\begin{equation}
u(k, t=0) = u_0(x)
\end{equation}
where $x=k_0 = k(t=0)$ (note that this $x$ is different from the one
used in section 3.2). Indeed one can note that 
for $u$ as a function of $k$ (which is what we want)
(\ref{char1}-\ref{char2}) leads to 
\begin{equation}
{\frac{du}{dk}} = {\frac{u'(t)}{k'(t)}} = 
-{\frac{1}{3}}{\frac{u}{k}}
\end{equation}
which can be integrated to:
\begin{equation}
\label{K1}
u = u_0 (x) (k/x)^{-1/3}
\end{equation}
This can now be used to solve 
(\ref{char1}-\ref{char2}) explicitly as:
\begin{equation}
\label{s1}
u(x,t)=u_0(x) (1 - 2 f(x) t)^{1/2}
\end{equation}
\begin{equation}
\label{s2}
k(x,t)=x (1 - 2 f(x) t)^{-3/2}
\end{equation}
where $f(x) = x u_0(x)$.

Two types of singularities occur in these solutions:
\begin{enumerate}
\item
$u(k)$ becomes multiple-valued when $k'(x) =0$. This will be called 
a ``turning point" and it implies the existence of a shock.
\item
Both $k(x)$ and $u(x)$ cease to exist beyond a certain finite time, 
where $k(x) \rightarrow \infty$. This will be referred to as the 
finite time singularity.
\end{enumerate}

For given initial conditions we can determine the singularities as 
function of time and thus the asymptotics of $u$. We assume that 
$u_0(x) >0$ (the opposite case will be treated in the next section), 
say monotonically increasing up to $x=x^*$ and monotonically 
decreasing for 
$x >x^*$. An example is 
\begin{equation}
\label{init}
u_0(x) = x e^{-x^2}
\end{equation}
 -- a localized initial disturbance. Note that the initial conditions can 
always be scaled by a constant, say $u = A v$ if time is also scaled as 
$t = T \tau$ such that
$T A = 1$.

The {\it turning time} $t=\alpha$ is found from
\begin{equation}
\label{alpha1}
k'(x) = (1-2 f(x) t )^{-5/2} (1 + (3 x f' - 2 f)t) = 0
\end{equation}
In particular the time $\alpha_1$ of the first turning point occurs at 
the minimal time for which
\begin{equation}
\label{alpha2}
1 + (3 x f'(x) - 2 f(x)) t  = 1+(3 x^2 u'_0 + x u_0) t=0
\end{equation}
and in the special case (\ref{init}) the minimum occurs when $2 - 8 
x^2 + 3 x^4 = 0$ or
\begin{equation}
\label{alpha3}
x_1 = \sqrt{(4 + \sqrt{10})/3} \approx 1.54
\end{equation}
($x = \sqrt{(4 - \sqrt{10})/3}$ must be discarded since $t<0$)
and
\begin{equation}
\label{alpha4}
\alpha_1 = {\frac{1}{2 f(x_1) (3 x_1^2 -2)}} \approx 0.44
\end{equation}

As $t$ increases we see from (\ref{alpha2}) that the turning point 
$x$ has
to occur in the regime where $u_0 <0$ and thus $x \geq x^*$. For 
the case (\ref{init}) we get as $t \rightarrow \infty$
we get $x_{\infty} = \sqrt{2/3} \approx 0.816$.

The {\it finite time singularity} occurs when $k(x) \rightarrow 
\infty$, which by (\ref{s2}) is equivalent to
\begin{equation}
\label{beta1}
t = \beta= {\frac{1}{2 f(x)}}
\end{equation}
The smallest value $\beta_1$ occurs at the maximum of $f(x)$. 
For the case (\ref{init}) this is $x=1$ and $\beta_1= e/2 \approx 
1.36$.

For very large times the solution makes sense up to
\begin{equation}
\label{beta3}
x_c (t) \approx (2t)^{-1/2}
\end{equation}
which is much smaller than $x_{\infty}\geq x^*$. At 
large times thus only the part $0 < x< x_c$ will contribute to the 
solution and the turning point becomes irrelevant. We can therefore use 
the 
form (\ref{K1}) expanded for small $x$ to find
\begin{equation}
\label{as1}
u(k,t) = x_c(t)^{4/3} k^{-1/3} \approx (2t)^{-2/3} k^{-1/3}
\end{equation}

\subsubsection{Shock conditions}
Due to the multiple valuedness this solution is cut off by a shock, 
where it jumps approximately to 0. The appropriate shock 
conditions 
have to be found by using the appropriate conservation law. It is 
reasonable to assume that this is the conservation of energy, 
which is the fundamental ingredient in the shell model.
In terms of the continuous wavenumber, the conservation of energy 
is
\begin{equation}
\label{e1}
{\frac{d}{dt}} \int u^2 {\frac{dk}{k}} =0
\end{equation}
which can be written as the local conservation law
\begin{equation}
\label{cons}
q_t = - j_k
\end{equation} 
i.e.
\begin{equation}
\label{e2}
({\frac{u^2}{k}})_t = - (2 u^3 k)_k
\end{equation}
As long as no shock occur (\ref{e2}) and (\ref{i-Parisi}) are 
equivalent, but crossing a shock the appropriate shock condition 
is \cite{Whitham}
\begin{equation}
\label{Rankine}
V = {\frac{d k}{dt}} ={\frac{[j]}{[q]}} 
\end{equation}
where $[ . ]$ denotes the discontinuity across the jump and $V$ is 
the velocity of the shock. Using energy conservation gives
\begin{equation}
\label{Rankine2}
V = {\frac{[2 u^3 k]}{[u^2 k^{-1}]}} =2 u k^2
\end{equation}

\subsubsection{Asymptotic forward pulse}
Combining the shock condition (\ref{Rankine2}) with the asymptotic 
form (\ref{as1}) gives for the edge of the shock $k_e(t)$
\begin{equation}
\label{as2}
{\frac{d k_e}{dt}} =2 (2t)^{-2/3}(k_e)^{5/3}
\end{equation}
which leads to
\begin{equation}
\label{as3}
k_e(t) = (A - c t^{1/3})^{-3/2}
\end{equation}
(where $c= 4\times 2^{-2/3}$ and $A$ is an unknown constant). Thus the 
shock position diverges at the finite time $t^* = (A/c)^3$ where the 
spectrum becomes of the Kolmogorov form 
all the way to the largest $k$ and only decays as $t^{-2/3}$.

\subsubsection{Backwards propagation}
In the backwards direction  we use the fact that
\begin{equation}
a={\rm Re}\, u = \sqrt{3} {\rm Im}\, u =\sqrt{3} b
\end{equation}
Then
\begin{equation}
b_t - 6 k^2 b b_k = 2 k b^2
\end{equation}
and now the substitution $u = -b/2$ gives us back (\ref{i-Parisi}) for 
$u$. Thus solving the propagation to smaller $k$ can be done by 
taking a {\it negative} initial condition for (\ref{i-Parisi}) and we 
thus have the solution as (\ref{s1}) -(\ref{s2})
with $u_0 < 0$, i.e. of the form
\begin{equation}
\label{init'}
u_0(x) = -x e^{-x^2}
\end{equation}

In this case the finite time singularity, where $k(x) \rightarrow 
\infty$ never occurs since $1-2 f t >0$ for all $x,t >0$. The turning 
points, which create shocks are again found by $k'(x) = 0$. The 
analysis is very similar to the forward case, but now, for the initial 
condition (\ref{init'}) the first shock 
appears at  $x_1 = \sqrt{(4 - \sqrt{10})/3} \approx 0.53$
with $\alpha_1\approx 2.04$.

For $t> \alpha_1$ the 
turning points occur at two distinct values of $x$, $x_a$ and 
$x_b > x_a$. For $t \rightarrow \infty$ we see from (\ref{alpha2}) 
that they are solutions of
$3 x^2 u'_0 + x u_0 =0$ which means that $x_a$ approaches the left 
hand edge of the interval of support for $u_0$ (which is zero for 
(\ref{init'})) and $x_b < x^*$ since $u'(x_b) <0$ (for 
(\ref{init'}), $x_b  \rightarrow \sqrt{2/3}$).
(It can be noted in passing that the maximum for $u$ also occurs 
very 
close to $x_b$, but at slightly larger $x$. The curve thus never 
crosses itself and only develops a cusp
in the limit $t \rightarrow \infty$.)

\subsubsection{Asymptotic left-going pulse}
Asymptotically the only relevant regimes of initial data are 
$0<x<x_a$ and $x_b<x<\infty$. For very small $x$ we simply 
expand in $x$ to get $u(k)$. The interesting regime is 
$x_b<x<\infty$ and here the main variation in $u$ comes from the 
regime where $-2 f t  \gg 1$. There
\begin{equation}
\label{as1'}
u k t = u_0 (1-2 f t)^{-3/2} x (1-2 f t)^{1/2} t = f t (1-2 f t)^{-1} 
\rightarrow -{\frac{1}{2}}
\end{equation}
or 
\begin{equation}
\label{as2'}
u(k,t)  \approx  -{\frac{1}{2 k t}}
\end{equation}
Again, the shock condition (\ref{Rankine2}) gives (since $u$ jumps 
to a very 
small value)
\begin{equation}
\label{as3'}
{\frac{d k_e}{dt}} =2 u k_e^2 = -{\frac{k_e}{t}}
\end{equation}
with the solution
\begin{equation}
\label{as4'}
 k_e(t) ={\frac{C}{t}}
\end{equation}
Asymptotically the solution  thus looks like (\ref{as2'})  down to
$k=k_e$ given by (\ref{as4'}) where it jumps to a small value 
(O(1/t)).
Note that the value of $u$ on the edge of the shock remains constant
\begin{equation}
\label{as5'}
 u_e \approx -{\frac{1}{2 C}}
\end{equation}
\section{Conclusion}
We have shown that an initial pulse on top of the zero solution 
of
the inviscid Parisi equations splits into two part, one for each
direction. The asymptotical solutions for the two pulses are:
\begin{eqnarray}
  u_{right} &=& (2t)^{-2/3}k^{-1/3} \\
  u_{left}  &=& (tk)^{-1}  \,\,\,\, {\rm down \,\, to \,\,} k\sim C/t
\end{eqnarray}
both decaying in time. Thus a burst created on top of the zero
solution not only travels down the inertial range until it is
dissipated by viscosity (here, by the continuum), it also travels 
upwards, to the smallest $k$. The burst does not remain localized in 
$k$-space, but is distributed over the $k$-range, and then decays. It
is interesting to note that the initial disturbance creates both a forward
and an inverse cascade.

We would like to thank Jakob Langgaard Nielsen for helpful 
discussions.

\end{document}